\documentstyle[twocolumn,epsf]{article}
\makeatletter
\renewcommand{\section}{%
\@startsection{section}{1}{\z@}
{8truept}{4truept}{\normalsize\bf}}
\renewcommand{\subsection}{%
\@startsection{subsection}{1}{\z@}
{8truept}{4truept}{\normalsize\bf}}
\def\fnum@figure{Fig.\hskip.5em\thefigure}
\long\def\@makecaption#1#2{
  \setbox\@tempboxa\hbox{{\bf #1}\quad #2}%
  \ifdim \wd\@tempboxa >\hsize \unhbox\@tempboxa\par
  \else \hbox to\hsize{\hfil\box\@tempboxa\hfil}
  \fi}
\makeatother
\begin{document}
\pagestyle{empty}
\title{\Large\bf Transport Properties in (Na,Ca)Co$_2$O$_4 $ Ceramics}
\author{\large T. Itoh$^1$, T. Kawata$^1$, T. Kitajima$^1$
and I. Terasaki$^{1,2,*}$\\
{\normalsize $^1$Department of Applied Physics, Waseda University,
Tokyo 169-8555, Japan}\\
{\normalsize $^2$Precursory Research for Embryonic Science and
Technology, Japan Science Technology Corporation}}
\date{}
\maketitle
\thispagestyle{empty}

\section*{Abstract}
The resistivity and thermopower of 
polycrystalline Na$_{1.1-x}$Ca$_x$Co$_2$O$_4$ were measured and analyzed.
Both the quantities increase with $x$, suggesting that the 
carrier density is decreased by the substitutions of Ca$^{2+}$
for Na$^{+}$.
Considering that the temperature dependence of the resistivity
show a characteristic change with $x$,
the conduction mechanism is unlikely to come from 
a simple electron-phonon scattering. 
As a reference for NaCo$_2$O$_4$, single crystals of 
a two-dimensional Co oxide Bi$_{2-x}$Pb$_xM_3$Co$_2$O$_9$ 
($M=$Sr and Ba) were studied.
The Pb substitution decreases the resistivity,
leaving the thermopower nearly intact.

\section{Introduction}
A search for new thermoelectric (TE) materials is an old problem
that has been reexamined to date \cite{Mahan}. 
Even though binary compounds might have been thoroughly studied,
a thermoelectric material of higher performance might sleep
in ternary, quaternary or more complicated compounds.
A filled skutterudite is an example for 
newly discovered TE materials \cite{Sales}.

Very recently Terasaki, Sasago and Uchinokura
have found that a layered Co oxide NaCo$_2$O$_4$,
whose crystal structure is schematically drawn in Fig. 1,
shows large thermopower (100 $\mu$V/K at 300 K)
and low resistivity (200 $\mu\Omega$cm at 300 K)
along the $a$ axis \cite{Tera}.
A striking feature of this compound is 
that the thermopower of 100 $\mu$V/K is realized in 
the carrier density of 10$^{21}$ cm$^{-3}$.
This is difficult to explain in the framework of
the conventional band picture,
and there should exist a mechanism 
to enhance the thermopower.

\begin{figure}[t]
\centerline{\epsfxsize=7cm 
\epsfbox{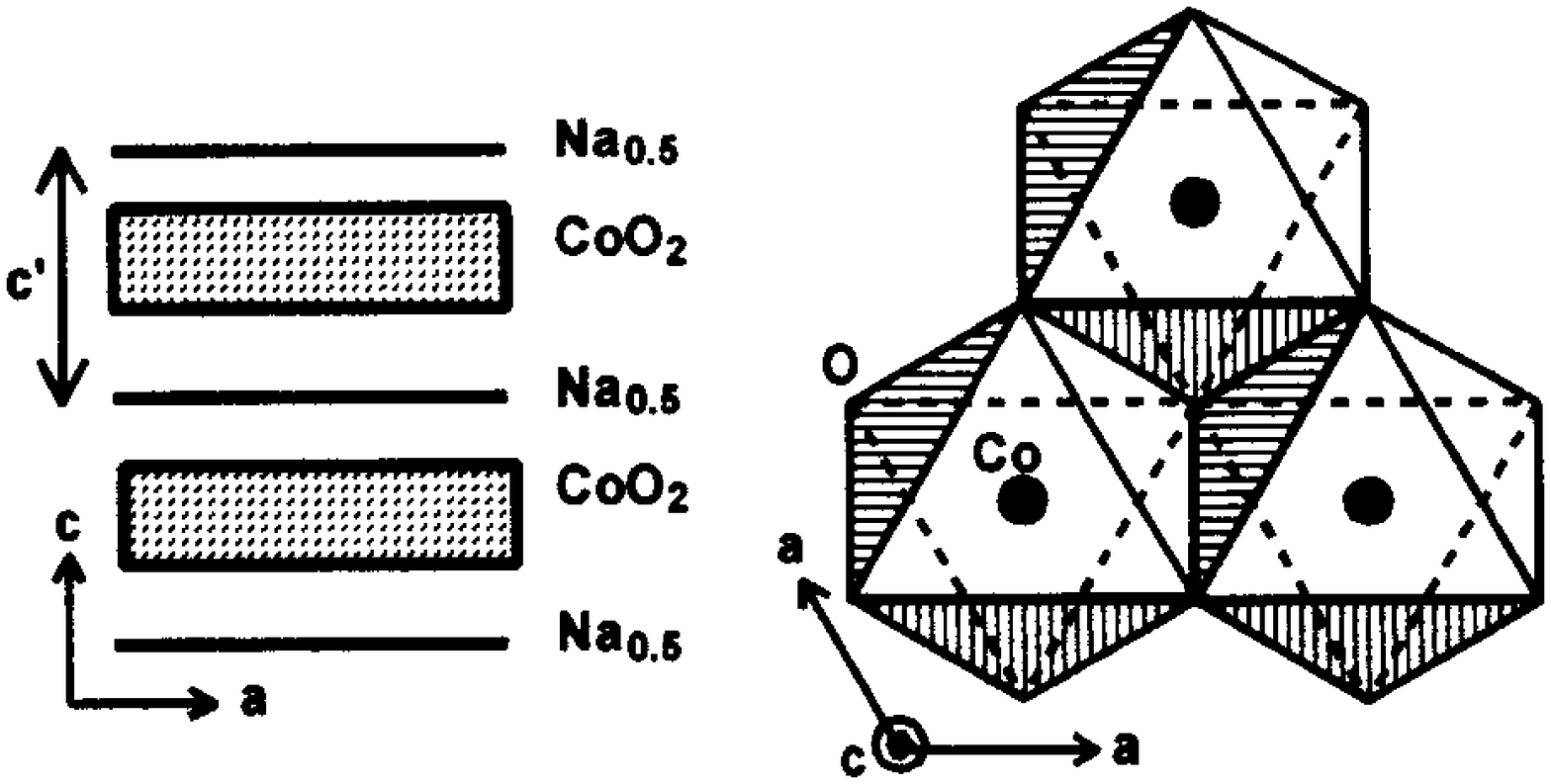}
}
\caption{Crystal structure  of NaCo$_2$O$_4$.}
\vspace*{1cm}
\centerline{\epsfxsize=5cm 
\epsfbox{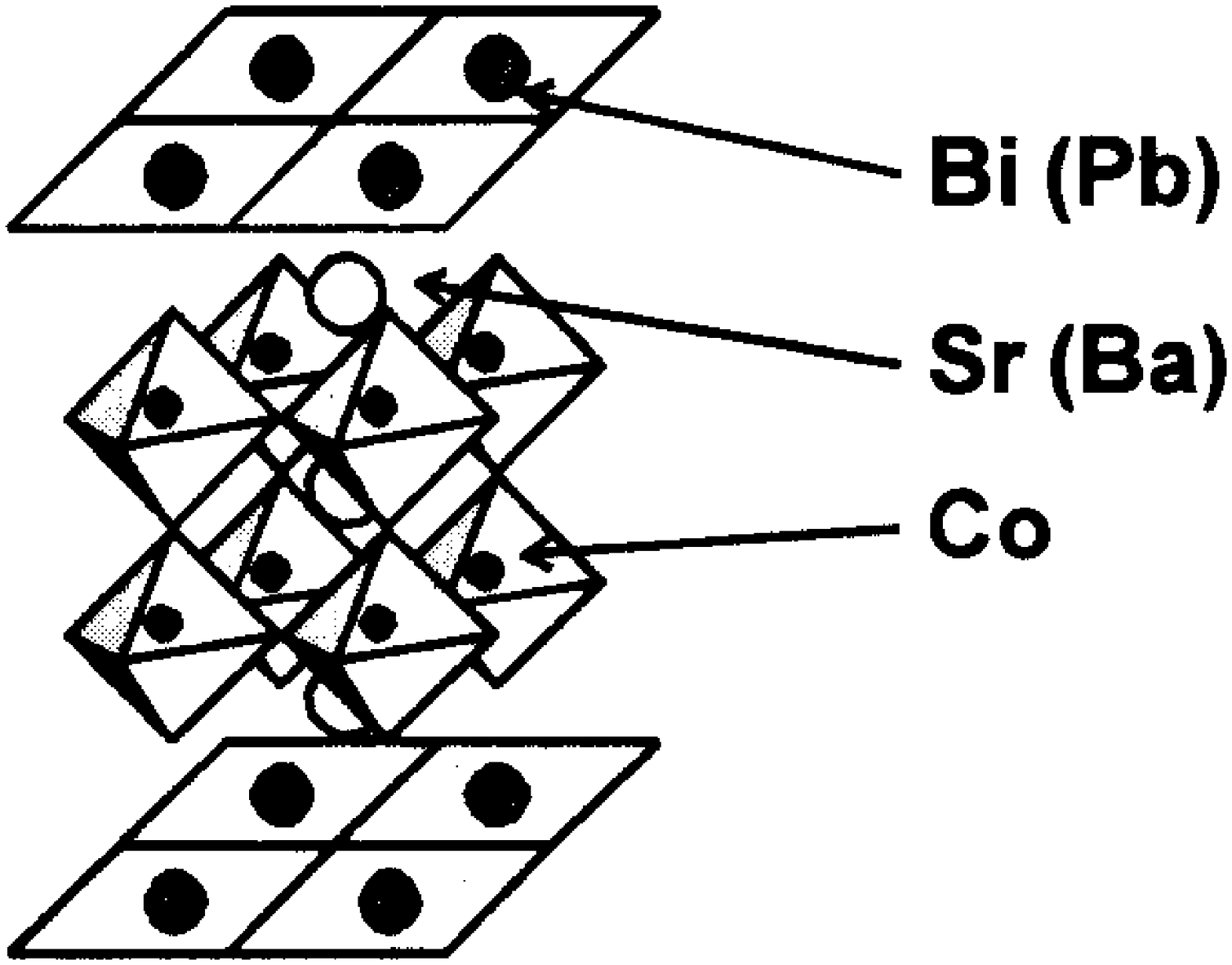}
}
\caption{ Crystal structure of Bi$_2M_3$Co$_2$O$_9$.}
\end{figure}

In conventional TE materials, the TE performance 
is optimized near a carrier density of $10^{19}$ cm$^{-3}$.
Thus we expect that the TE performance of NaCo$_2$O$_4$
may be improved by the reduction of the carrier density.
The easiest way to change the carrier density is to
substitute a divalent cation such as Ca$^{2+}$ for a monovalent Na$^+$.   
Motivated by this, we measured and analyzed the resistivity and thermopower 
of polycrystalline (Na,Ca)Co$_2$O$_4$.
Another way is to study a two-dimensional Co oxide with low carrier density.
For this purpose Bi$_2$Sr$_3$Co$_2$O$_9$ 
(The crystal structure is shown in Fig. 2) is most suitable,
because the resistivity is lowered by the Ba-substitution for Sr \cite{Tara}
and the Pb substitution for Bi \cite{Tsuka}.
In addition, the optical reflectivity shows a small Drude weight \cite{Tera2}.  
We report on its transport properties in the 
latter part of the proceedings.

\section{Experimental}
Polycrystalline samples of Na$_{1.1-x}$Ca$_x$Co$_2$O$_4$ were
prepared in a conventional solid-state reaction \cite{Yakabe}.
An appropriate mixture of powdered NaCO$_3$, CaCO$_3$ and Co$_3$O$_4$
was calcined at 860$^{\circ}$C for 12 h.
The product was finely ground, pressed into a pellet,
and sintered at 800$^{\circ}$C for 6 h.
The x-ray diffraction pattern showed no trace of impurities.
Note that the a tiny trace of impurity phases
was detected in the product from stoichiometric mixture (Na:Co=1:2),
which indicates the evaporation of a small amount of Na.
Thus we added excess Na of 10 at.\% to prepare NaCo$_2$O$_4$.

Single crystals of Bi$_2M_3$Co$_2$O$_9$ ($M$=Sr and Ba) \cite{Tara} and
Bi$_{2-x}$Pb$_x$Sr$_3$Co$_2$O$_9$ \cite{Tsuka}
were prepared by a self-flux technique.
Crystals were platelike with typical dimensions of 
1$\times$1$\times$0.01 mm$^3$.
As for the Pb substitution,
we prepared two different crystals 
with nominal compositions of $x$=0.2 and 0.4.

Resistivity ($\rho$) was measured through a four-probe method.
Thermopower ($S$) was measured with a nano-voltmeter (HP 34420A),
where a typical resolution was 5--10 nV.
Two edges of a sample was pasted on Cu sheets working as a heat bath, 
and the temperature gradient of 0.5--1 K was measured 
through a differential thermocouple made of copper-constantan.
The contributions from copper leads were carefully subtracted. 
 
\begin{figure}[t]
\centerline{\epsfxsize=7cm 
\epsfbox{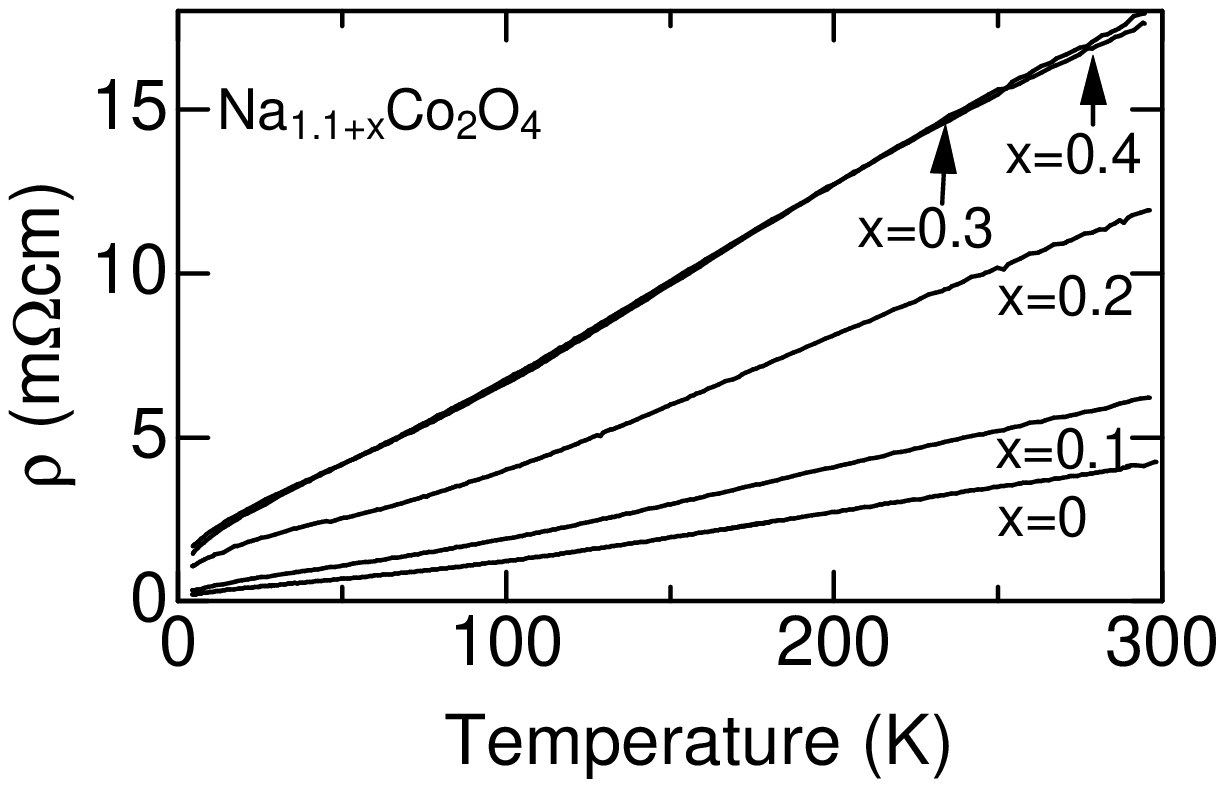}
}
\caption{Resistivity
of polycrystalline samples of Na$_{1.1+x}$Co$_2$O$_4$.}
\vspace*{1cm}
\centerline{\epsfxsize=7cm 
\epsfbox{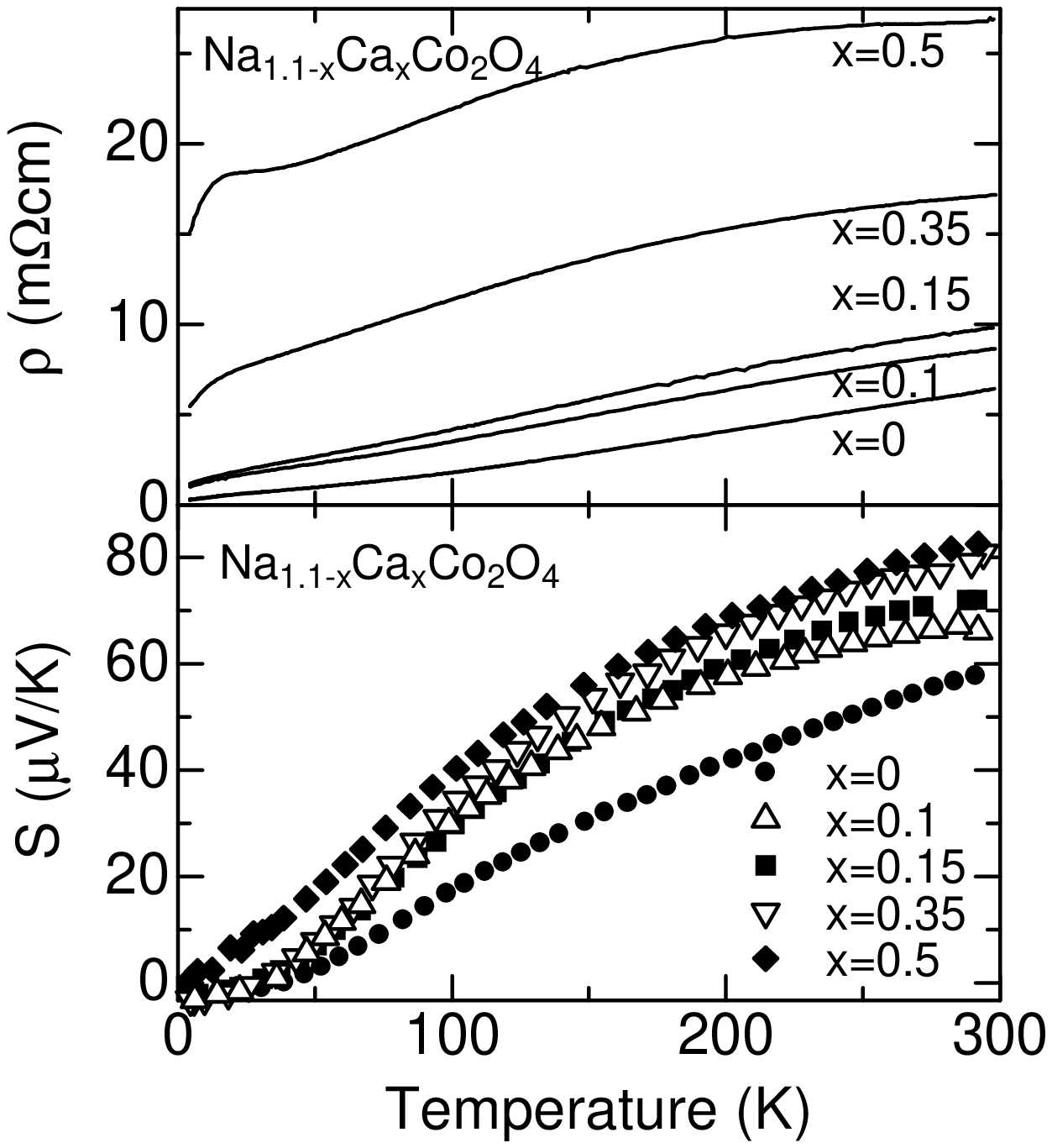}
}
\caption{(a) Resistivity and (b) thermopower 
of polycrystalline samples of Na$_{1.1-x}$Ca$_x$Co$_2$O$_4$.}
\end{figure}

\section{Results and Discussion}
Let us begin with the effect of excess Na on 
the resistivity of Na$_{1.1+x}$Co$_2$O$_4$.
Na is an element difficult to control.
First, it is quite volatile above 800$^{\circ}$C.
Secondly, residual Na is rarely observed in x-ray diffraction patterns,
because it often exists as deliquesced NaOH in the grain boundary.
Moreover the Na site in NaCo$_2$O$_4$ is 50\% vacant,
that is, the Na content $x$ can change from 0 to 1.
Figure 3 shows the temperature dependence of $\rho$
of polycrystalline Na$_{1.1+x}$Co$_2$O$_4$.
We attributed the increase of $\rho$ with $x$
to the excess Na in the grain boundaries,
and regarded the sample for $x=0$ as the parent material
for Ca substitution.

Figures 4(a) and 4(b) show $\rho$ and $S$ of polycrystalline 
samples of Na$_{1.1-x}$Ca$_x$Co$_2$O$_4$.
The magnitudes of $\rho$ and $S$ increase with $x$,
suggesting that the carrier density is reduced by the Ca substitution.
This is naturally understood from a viewpoint of Co valence.
In NaCo$_2$O$_4$ the formal valence ($p$) of Co is $3.5+$,
{\it i.e.}, Co$^{3+}$:Co$^{4+}$=1:1.
The Ca substitution decreases $p$ down to $3+$,
which corresponds to the configuration of $(3d)^6$.
Since the six electrons fully occupied the three $d\gamma$ bands
in the low spin state,
oxides with Co$^{3+}$ are often insulating.
As expected, the power factor $S^2/\rho$ is improved in $x\sim$0.15 by 20\%.
It should be emphasized that the Ca substitution changes 
the temperature dependence of $\rho$.
For example, while $\rho$ for $x$=0 shows a positive curvature 
below 100 K, $\rho$ for $x$=0.35 shows a negative curvature.
This indicates that the scattering rate depends strongly on the
carrier density, which is unlikely to
arise from the electron-phonon scattering.

\begin{figure}[t]
\centerline{\epsfxsize=7cm 
\epsfbox{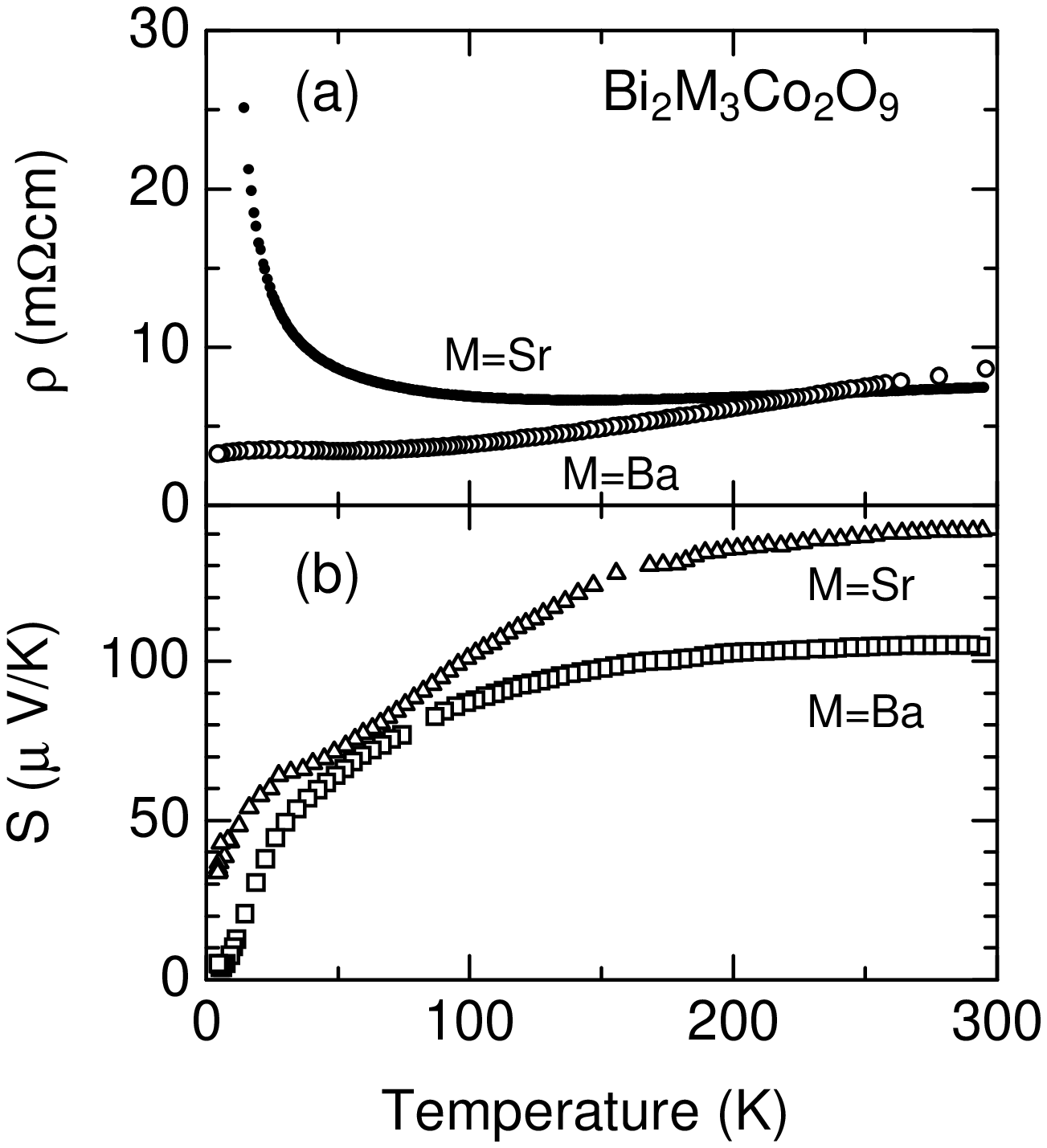}
}
\caption{(a) In-plane resistivity and (b) in-plane thermopower 
of single crystals of 
Bi$_2M_3$Co$_2$O$_9$ ($M$=Sr and Ba).}
\vspace*{1cm}
\centerline{\epsfxsize=7cm 
\epsfbox{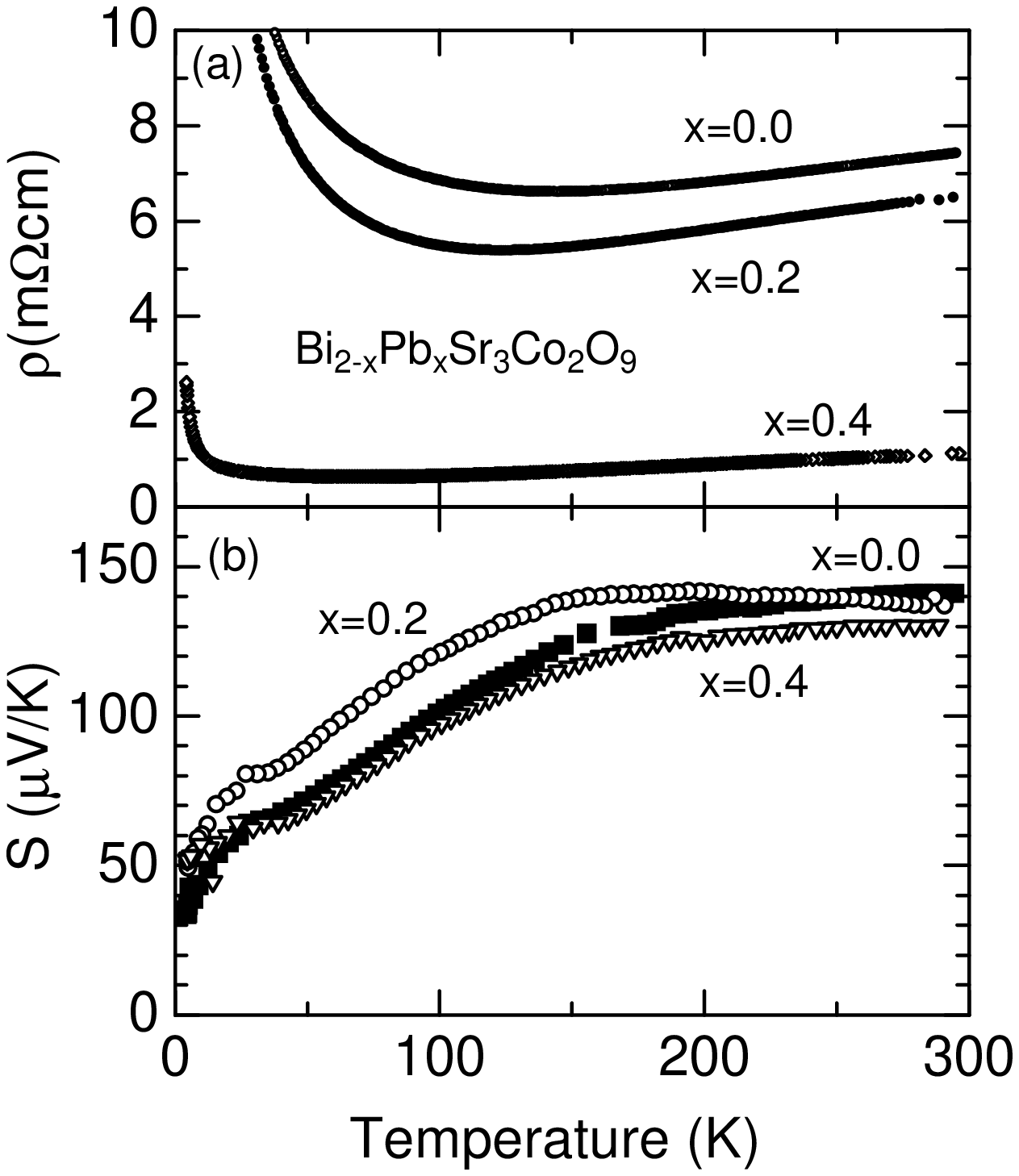}
}
\caption{(a) In-plane resistivity and (b) in-plane thermopower 
of single crystals of 
Bi$_{2-x}$Pb$_x$Sr$_3$Co$_2$O$_9$.}
\end{figure}

Next we discuss the thermoeletric properties of 
(Bi, Pb)$_2M_3$Co$_2$O$_9$.
In Figs. 5(a) and 5(b), $\rho$ and $S$ of
Bi$_2M_3$Co$_2$O$_9$ single crystals along the in-plane direction
are plotted as a function of temperature.
$\rho$ of the present samples reproduces the data in the literature,
where the electric conduction for $M$=Ba is more
metallic than that for $M$=Ba \cite{Tara}.
Note that the magnitude of $S$ is above 100 $\mu$V/K for both samples,
owing to the small carrier density.

In contrast to the Ba substitution for Sr,
Pb not only works as an acceptor, but also modifies
the electronic states of Bi$_2$Sr$_3$Co$_2$O$_9$.
Figures 6(a) and 6(b) show $\rho$ and $S$ of
Bi$_{2-x}$Pb$_x$Sr$_3$Co$_2$O$_9$ single crystals along the
in-plane direction.
$\rho$ is decreased drastically by the Pb substitution,
which looks similar to Fig. 5(a).
However, $S$ remains nearly unchanged upon the Pb substitution,
which clearly indicates that the Ba and Pb substitutions
affect the electronic states differently.
The anomalous electronic states of Bi$_{2-x}$Pb$_x$Sr$_3$Co$_2$O$_9$
are also suggested by the large negative magnetoresistance at low
temperatures \cite{Tsuka}.

\section{Summary}
In summary, we prepared polycrystalline samples 
of Na$_{1.1-x}$Ca$_x$Co$_2$O$_4$
and single-crystal samples of Bi$_{2-x}$Pb$_xM_3$\-Co$_2$O$_9$ 
($M$= Sr and Ba; $x$=0, 0.2 and 0.4).
In Na$_{1.1-x}$Ca$_x$Co$_2$O$_4$, both the resistivity and the
thermopower increase with $x$, which suggests that the
carrier density is decreased by the Ca substitution.
In Bi$_{2-x}$Pb$_xM_3$Co$_2$O$_9$, the resistivity 
is decreased by the Ba and Pb substitutions,
but the doping effects on the thermopower are different.
While the Ba substitution decreases the thermopower,
the Pb substitution hardly changes the thermopower.
This is a direct example that the resistivity can be lowered
while the thermopower is kept large.

\section*{Acknowledgements}
The authors would like to thank M. Takano, S. Nakamura, K. Fukuda,
S. Kurihara and K. Kohn for fruitful discussions.
They also appreciate H. Yakabe, K. Nakamura, K. Fujita and K. Kikuchi 
for collaboration.
They are indebted to I. Tsukada for showing us 
the unpublished data of (Bi,Pb)$_2$Sr$_3$Co$_2$O$_9$.


\begin{thebibliography}{99}
\bibitem[*]{byline}
Corresponding author: Ichiro Terasaki\\
\hspace*{8mm}Phone and Fax: +81-3-5286-3854\\
\hspace*{8mm}E-mail: terra@mn.waseda.ac.jp.

\bibitem{Mahan} 
G. Mahan, B. Sales and J. Sharp,
``Thermoelectric materials: New approaches to an old problem'',
Physics Today, March, pp. 42-47 (1997).

\bibitem{Sales}
B. C. Sales, D. Mandrus, B. C. Chakoumakos, V. Keppens
and J. R. Thompson,
``Filled skutterudite antimonides: 
Electron crystals and phonon glasses'',
Phys. Rev. B, Vol. 56, No. 23, pp. 15081-15089 (1997).

\bibitem{Tera}
I. Terasaki, Y. Sasago and K. Uchinokura,
``Large thermoelectric power of NaCo$_2$O$_4$ single crystals'',
Phys. Rev. B, Vol. 56, No. 20, pp. R12685-R12687 (1997).


\bibitem{Tara}
J. M. Tarascon, R. Ramesh, P. Barboux, M. S. Hedge, G. W. Hull, 
L. H. Green, M. Giroud, Y. LePage, W. R. McKinnon, J. V. Waszczak,
and L. F. Schneemeyer, 
``New non-superconducting layered Bi-oxide phases of formula 
Bi$_2M_3$Co$_2$O$_y$ containing Co instead of Cu'',
Solid State Commun., Vol 71, No. 8, pp. 663-668 (1989).


\bibitem{Tsuka}
I. Tsukada, T. Yamamoto, M. Takagi, T. Tsubone and 
K. Uchinokura,
``Negative magnetoresistance in  (Bi,Pb)$_2$Sr$_3$Co$_2$O$_9$
layered cobalt oxides'',
Mat. Res. Soc. Symp. Proc., Vol.494, pp. 119-125 (1998). 

\bibitem{Tera2}
I. Terasaki, T. Nakahashi, A. Maeda and K. Uchinokura
``Optical reflectivity of single-crystal Bi$_2M_3$Co$_2$O$_9$ 
($M$=Ca, Sr and Ba) from infrared to vacuum-ultraviolet region'',
Phys. Rev. B, Vol. 47, No. 1,  pp. 451-456 (1993)

\bibitem{Yakabe}
H. Yakabe, K. Kikuchi, I. Terasaki, Y. Sasago and K. Uchinokura,
``Thermoelectric propoerties of transition-metal oxide 
NaCo$_2$O$_4$ system'',
in Proceedings of the XVIth
International Conference on Thermoelectrics,
Dresden, Germany, August 26-29, 1997 (in press).

\bibitem{HF} 
For example, C. S. Garde and J. Ray, 
``Thermopower and resistivity behavior in Ce-based Kondo-lattice systems:
A phenomenological approach'',
Phys. Rev. B, Vol. 51, No. 5, pp. 2960-2965 (1995).


\end{thebibliography}
\end{document}